\documentclass[11pt]{article}
\pdfoutput=1
\usepackage{graphicx,amssymb,amsfonts,amsmath,amssymb,amscd,amstext, mathrsfs}
\usepackage{color}
\makeatletter

\@addtoreset{equation}{section}
\newcommand{\be}{\begin{eqnarray}}
\newcommand{\ee}{\end{eqnarray}}
\newcommand{\ba}{\begin{array}}
\newcommand{\ea}{\end{array}}
\newcommand{\nn}{\nonumber}

\makeatletter \@addtoreset{equation}{section} \makeatother

\begin{document}
\vspace{1cm}
\begin{center}
~\\~\\~\\
{\bf  \LARGE Operator Approach to Complexity : Excited States}
\vspace{1cm}

                      Wung-Hong Huang\\
                       Department of Physics\\
                       National Cheng Kung University\\
                       Tainan, Taiwan\\

\end{center}
\vspace{1cm}
\begin{center}{\bf  \Large ABSTRACT}\end{center}
We evaluate the complexity of the  free scalar field by  the operator approach in which the transformation matrix between the second quantization operators of reference state and target state is regarded as the quantum gate. We first  examine the system in which the reference state is  two non-interacting oscillators with same frequency $\omega_0$ while the  target state  is  two interacting oscillators with frequency $\tilde \omega_1$ and $\tilde \omega_2$.   We calculate the geodesic length on the associated  group manifold of gate matrix and    reproduce the known value of ground-state complexity. Next,  we study the complexity in the excited states.   Although the  gate  matrix is very large we can transform  it to a diagonal matrix and obtain the  associated complexity.  We explicitly calculate the complexity in several excited states  and  prove that the square of geodesic length in the general state $|{\rm n,m}\rangle$ is $D_{\rm (n,m)}^2={\rm (n+1)}\left(\ln {\sqrt{\tilde \omega_1\over  \omega_0}}\,\right)^2 +{\rm (m+1)}\left(\ln {\sqrt{\tilde \omega_2\over  \omega_0}}\,\right)^2$.  The results are extended to the N couple harmonic oscillators which correspond to  the lattice version of  free scalar field.
\vspace{2cm}
\begin{flushleft}
*E-mail:  whhwung@mail.ncku.edu.tw\\
\end{flushleft}
\newpage
\tableofcontents
\section{Introduction}
Complexity plays an important role  in understanding how spacetime emerges from field theory degrees of freedom within the AdS/CFT correspondence \cite{Raamsdonk1005, Swingle1405, Lashkari1308, Faulkner1312}, besides the entanglement entropy \cite{Ryu0603} .  It relates to  the tensor network models and  involves the dynamics of black hole interiors \cite{Maldacena0106}.  In the context of the eternal AdS-Schwarzchild black hole the wormhole which connects the two sides grows linearly with time  \cite {Hartman1305}, which is  conjectured to dual to the growth of complexity of the dual CFT state \cite { Carmi1709, Ali1810,  Ali1811}.

In the context of AdS/CFT two interesting proposals are used  to evaluate the holographic complexity.  The first is complexity=volume (CV) conjecture \cite{Susskind1403} which  suggests that complexity is dual to the volume of an extremal (codimension one) bulk surface anchored to a certain time slice in the boundary. The second is  complexity=action (CA) conjecture  \cite{Brown1509, Brown1512, Chapman1701} which identifies the complexity with the gravitational action evaluated on a particular bulk region, known as the Wheeler-DeWitt (WDW) patch. 

Complexity is the number of  operations $\{\cal O^I\}$ needed to transform a reference state $|\psi_R\rangle$ to a target state $|\psi_T\rangle$. The operators are called as quantum gates  and the more gates we need the more complex the target state was. We can define the  affine parameter ``s" associated to an unitary operator $U(s)$ and  use a set of function $Y^I(s)$ to character the quantum circuit. The unitary operation which connects the reference state and target state is
\be
U(s)=\vec{\cal P}\, e^{-\int_0^s\,Y^I(s)\,{\cal O}^I},~~|\psi_R\rangle =U(0)|\psi_R\rangle,~~~|\psi_T\rangle =U(1)|\psi_R\rangle~~~\label{O}
\ee
where $\vec{\cal P}$ indicates a time ordering  along $s$. The circuit depth $D[U]$ which also called as cost function is defined by
\be
D[U]&=&\int_0^1ds\,\sqrt{\sum_I |Y^I(s)|^2}
\ee
Use it the  complexity ${\cal C}$ is defined by
\be
{\cal C}&=&\underset{\rm \{Y^I\}}{\rm Min}\,D[U]
\ee
For the models which had been studied the quantum gates are the group generators.  Thus the minimum in ${\cal C}$   means that we  are to calculate the geodesic in the Riemann space of group manifold.

The calculation along this line was initial in papers  \cite{Jefferson1707}  and  \cite{Chapman1707} which considered the free scalar field.  Later,  it was extended to study the free fermions \cite{Khan1801, Hackl1803,Jiang1812}, quenched system \cite{Alves1804, Camargo1807, Jiang1810}, coherence state \cite{Guo1807}, and interacting model \cite{Bhattacharyya1880}.  The investigate in  \cite{Jefferson1707} had shown that once the cost function is chosen to be
\be
D[U]&=&\int_0^1ds\,\sum_I |Y^I(s)|^2
\ee
the field theory calculation could match to gravity method.

The most studies in field theory are considering the Gaussian ground state \cite{Jefferson1707,Khan1801, Hackl1803} in the free field or exponential type  wavefunction in interacting model \cite{Bhattacharyya1880}, which are reviewed in next section. 

 In this paper we will present  the  operator approach to evaluate the complexity in free scalar field theory. {\color{black}Note that the operator approach had been used in \cite{Chapman1707, Hackl1803,Alves1804, Camargo1807} which study the  fermion theory and quench mode.  In these papers the reference operator and target operator are the creation operators with different mass or frequency. The Bogoliubov transformations  between the operators relate to the quantum gate of the theory. In our study the reference operator is the creation operator of non-interacting harmonic oscillator while the target operator is that in the  interacting harmonic oscillator. We  regard the transformation between the create operators of reference state  and target state as the quantum gate and calculate the geodesic length on the associated group manifold.  Since that in the  operator approach we need not to use the explicit form of the wave function we can study the complexity in the excited states. Note that the excited-state wavefunction of harmonic oscillation  is not pure  exponential form and  the wavefunction approach is hard to work, as commented in section 2.}\\

The paper is organized as follows. In section 2,  we quickly review the method in  \cite{Jefferson1707}. After  describing the lattice free scalar field as coupled harmonic oscillators we consider first the 2 coupled oscillators  system.  We review how the transformation between the ground-state wavefunction of the reference state (i.e. two non-interacting oscillators) and target state (i.e. two interacting oscillators)  can be used to calculate the complexity of the ground state.  In section 3, we setup the notation and describe  the transformation  between the second quantization operators  in the reference state and target state.  We find that the gate matrix  can be reduced to diagonal form after orthogonal transformation.  Then, we calculate circuit depth $D[U]$ and the complexity. The result reproduces the known value of ground-state complexity in \cite{Jefferson1707}. 

In section 4  we carefully calculate the complexity in several  excited state and  prove that the square of geodesic length in the general state $|{\rm n,m}\rangle$ is $D_{\rm (n,m)}^2={\rm (n+1)}\left(\ln {\sqrt{\tilde \omega_1\over  \omega_0}}\,\right)^2 +{\rm (m+1)}\left(\ln {\sqrt{\tilde \omega_2\over  \omega_0}}\,\right)^2$.   In section 5   we describe how the calculations in previous sections, which are focused  on two coupled harmonic oscillators, can be extended to N coupled harmonic oscillators which describes lattice free scalar field theory.  Final summary appears in the last section. In appendix A we discuss the relation of geodesic length between a  gate matrix before and after transformed to a diagonal form.   In appendix B we derive the  summation formulas which are used to derive the complexity formula in the  excited states.
\section{Free Scalar Field and Harmonic Oscillator : Wavefunction Approach}
We consider the Hamiltonian of a free scalar field in $d$ spacetime dimensions 
\be
H=\frac{1}{2}\int d^{d-1}x\left[\pi(x)^2+\vec\nabla\phi(x)^2+m^2\phi(x)^2\right]~.
\ee
Placing the theory on a square  lattice with lattice spacing  $\delta$  the Hamiltonian is described by a lattice version  \cite{Jefferson1707}
\be
H=\frac{1}{2}\sum_{\vec n}\left\{\frac{p(\vec n)^2}{\delta^{d-1}}+\delta^{d-1}\Big[\frac{1}{\delta^2}\sum_i\Big((\phi(\vec n)-\phi(\vec n-\hat{x}_i))^2+m^2\phi(\vec n)^2\Big]\right\}
\ee
where $\hat{x}_i$ are unit vectors pointing along the spatial directions of the lattice. By redefining $X(\vec n)=\delta^{d/2}\phi(\vec n)$, $P(\vec n)=p(\vec n)/\delta^{d/2}$, $M=1/\delta$, $\omega=m$ and $\Omega=1/\delta$  the lattice Hamiltonian  becomes
\be
H=\sum_{\vec n}\left\{\frac{P(\vec n)^2}{2M}+\frac12 M \left[\omega^2 X(\vec n)^2+\Omega^2\sum_i\Big( X(\vec n)-X(\vec n-\hat{x}_i)\Big)^2\right]\right\}~~~\label{LH}
\ee
where $\omega=m$ and  $\Omega=1/\delta$.  {\color{black}The resulting theory of free scalar field is essentially a quantum mechanical problem with an infinite family of coupled  harmonic oscillators}. 

Consider first a simple case of two coupled harmonic oscillators:
\be
H=\frac{1}{2}\left[ p_1^2+p_2^2+\omega^2( x_1^2+x_2^2)+\Omega^2(x_1-x_2)^2\right]~~~\label{LH1}
\ee
where $x_1,x_2$ label their spatial positions, after setting $M_1=M_2=1$ for simplicity.  The Hamiltonian expressed in terms of the normal modes is 
\be
H=\frac{1}{2}\Big( \tilde p_1^2+\tilde\omega_1^2\tilde x_1^2+\tilde p_2^2+\tilde\omega_2^2\tilde x_2^2\Big)
\ee
where
\be
\tilde x_{\{1,2\}}\equiv\frac{1}{\sqrt{2}}\left( x_1\pm x_2\right),\qquad\tilde p_{\{1,2\}}\equiv\frac{1}{\sqrt{2}}\left(p_1\pm p_2\right),\qquad
\{\tilde\omega_1^2,\tilde\omega_2^2\}=\{\omega^2,\omega^2+2\Omega^2\}
\ee
The normalized ground-state wave function, which chosen as the target state,  becomes 
\be
\psi_T=\psi_{0}(\tilde x_1)\psi_{0}(\tilde x_2)
=\frac{\Big(\tilde\omega_1\tilde\omega_2\Big)^{1/4}}{\sqrt{\pi}}\,\mathrm{exp}\!\left[-\frac{1}{2}\Big(\tilde\omega_1\tilde x_1^2+\tilde\omega_2\tilde x_2^2\Big)\right]~,\label{eq:targetNorm}
\ee
We can express this wave function in terms of the physical positions of the two masses:
\be
\psi_T=\frac{\Big(\omega_1\omega_2-\beta^2\Big)^{1/4}}{\sqrt{\pi}}\mathrm{exp}\left[-\frac{\omega_1}{2}x_1^2-\frac{\omega_2}{2}x_2^2-\beta x_1x_2\right],\label{target}
\ee
where
\be
\omega_1=\omega_2=\frac{1}{2}\Big(\tilde\omega_1+\tilde\omega_2\Big)~,\;\;\;
\beta\equiv\frac{1}{2}\Big(\tilde\omega_1-\tilde\omega_2\Big)
\ee
Above Gaussian wave functions is the target state. The reference state, as that in \cite{Jefferson1707},  is chosen to be following factorized Gaussian state in which the two masses are unentangled
\be
\psi_{R}=\sqrt{\frac{\omega_0}{\pi}}\,\mathrm{exp}\!\left[-\frac{\omega_0}{2}\Big( x_1^2+x_2^2\Big)\right]~\label{ref}
\ee
where $\omega_0$ is a free parameter which characterizes our reference state.  {\color{black} This means that we will compute the complexity of the interacting ground states relative to  the non-interacting ground states}.

After choosing the reference and target states we have to find a  unitary transformation $U$ which implements $\psi_{T}=U\,\psi_{R}$.  In the wavefunction approach the transformation is considered to  relate the the exponential part in target state wavefunction (\ref{target}) and in reference state wavefunction (\ref{ref}), i.e.
\be
\omega_0\,\left( x_1^2+x_2^2\right)\rightarrow \omega_1x_1^2+\omega_2x_2^2+2\beta x_1x_2
\ee
 We can use the basic vector $\psi_{R}^T=\left(\sqrt {\omega_0}\,x_1,\sqrt {\omega_0}\,x_2\right)$ to express above transformation in a matrix form
\be
\psi_{R}^T\,\cdot\,\psi_{R}\rightarrow \left(\psi_{R}^TU^T\right)\,\cdot\,\left(U\psi_{R}\right)=\psi_{R}^T\,\cdot\,{\rm M_ w}\,\cdot\,
\,\psi_{R}~~\label{eq1},~~
{\rm M_ w}=U^TU=\left(\ba{cc}{\omega_1\over \omega_0}&{\beta\over \omega_0}\\
{\beta\over \omega_0}&{\omega_2\over \omega_0}\ea\right)~~~~\label{Mw}
\ee
in which ${\rm M_ w}$ denotes the transformation matrix in the wavefunction approach.

 {\color{black}Gate matrix $U(s)$ belongs to group $GL(2,R)=R\times SL(2,R)$ which can be expressed as a \cite{Jefferson1707}}
\be
U_{GL(2,R)}(s)=e^{y(s)}\!\left(\!\!\!\ba{cc}
\cos\tau(s)\cosh\rho(s)\!-\!\sin\theta(s)\sinh\rho(s)&-\!\sin\tau(s)\cosh\rho(s)\!+\!\cos\theta(s)\sinh\rho(s)\\
\sin\tau(s)\cosh\rho(s)\!+\!\cos\theta(s)\sinh\rho(s)&\cos\tau(s)\cosh\rho(s)\!+\!\sin\theta(s)\sinh\rho(s)
\ea\!\!\!\right)~\label{U}~\label{GL}\nn\\
\ee
{\color{black} Regard the operator ${\cal O}^I$ in  (\ref{O}) as a group generators of matrix  $M^I$  the equation (\ref{O}) has solution
\be
Y^I(s)={1\over {\rm Tr} \left((M^J)^T\,M^J\right)}{\rm Tr}\left( \partial_sU(s)U(s)^{-1}(M^I)^T\right)
\ee
 The line element on the group  manifold  becomes 
\be
ds^2&=&\delta_{IJ}dY^I(s)dY^I(s)\nn\\
&=&2\Big(dy^2+d\rho^2+\cosh(2\rho)\cosh^2\rho d\tau^2+\cosh(2\rho)\sinh^2\rho d\theta^2-\sinh^2(2\rho)d\tau d\theta\Big)
\ee
in which the metric $g_{IJ}=\delta_{IJ}$ is that chosen in \cite{Jefferson1707}. The geodesics on above Riemann space had been analyzed and it found that the square of geodesic length becomes  
\be
Distance^2&=&\int_0^1ds\,\,\delta_{IJ}\,{dY^I(s)\over ds}{dY^I(s)\over ds}\nn\\
&=&2y(1)+2\rho(1)~~~~\label{0}
\ee
Note that the geodesic solution is that with initial condition $\left((y(0),\rho(0),\tau(0),\theta(0)\right)=(0,0,0,\theta_0)$.  While $\theta_0$ is undetermined dues to the rotation symmetry in the group manifold we can choose $\theta(0)=\theta(1)=\theta_0=0$ in general. }{\color{black} Also, the solution found in \cite{Jefferson1707} shows that $\tau(s)$ is constant along the geodesic trajectory. Thus $\tau(1)=\tau(0)=1$ and it does not contribute to  the geodesic length. The other parts of geodesic trajectory solutions are  $y(s)=y(1)\cdot s$ and $\rho(s)=\rho(1)\cdot s$.}

Use (\ref{Mw}) the solution of $U^T(1)U(1)={\rm M_ w}$  is
\be
y(1)&=&{1\over 4}\ln{\tilde \omega_1\tilde \omega_2\over \omega_0^2},~~~~~\rho(1)={1\over 4}\ln{\tilde \omega_2\over \tilde \omega_1},~~~~~~~~~~~\label{trho}
\ee
thus, the {\color{black} geodesic distance in the  manifold of Riemann space of group $GL(2,R)$ is}
\be
D^2=2(y(1))^2+\rho(1)^2)={1\over 4}\Big((\ln {\tilde \omega_1\over \omega_0})^2+(\ln{\tilde \omega_2\over \omega_0})^2\Big)=\left(\ln \alpha\right)^2+\left(\ln\gamma\right)^2~~~~~\label{Distance}
\ee
Above is that described by Jefferson and Myers in \cite{Jefferson1707}. 

  The method  relies on the Gaussian type wavefunctions in (\ref{target}) and (\ref{ref}) in which the quantum gate is an $GL(2,R)$ matrix in (\ref{GL}) and the basic vectors are $\left(\sqrt {\omega_0}\,x_1,\sqrt {\omega_0}\,x_2\right)$ The gate matrix is used to transform the exponential part of the wavefunctions from the reference state to the target state. 

The method had been  extended to study $\lambda\,\Phi^4$ \cite{Bhattacharyya1880}. In the leading order of small $\lambda$ the ground state wavefunctions of target state and reference state can be expressed as the exponential type. For example the reference state wavefunction  is $\Psi\sim exp\left[-{\omega\over 2}\left(x_1^2+x_2^2+\lambda(x_1^4+x_2^4+x_1^2x_2^2)\right)\right]$. Now the  basic vectors are $(x_1, x_2, x_1x_2,x_1^2, x_2^2)$ and quantum gate becomes  a $GL(5,R)$ matrix.   While the wavefunction becomes  more complex it can be studied along the free scalar field case since the wavefunction is pure exponential type.

In the cases of the excited harmonic states the wavefunctions  are not described by the  pure exponential form.  For example the wave function of the first excited state of harmonic state  is $x \, e^{-{1\over 2}\omega x^2}$ which is not a pure exponential form, and then it is hard to work in the wavefunction approach. 

In the following sections we will turn to the operator approach which needs not the explicit function form of wavefunction and can be used to study the complexity in the excited states.
\section{Complexity in Ground State of 2 Harmonic Oscillators}
\subsection{Basic Scheme}
In the second quantization the target ground state $|0,0\rangle_{\rm target}$ and the reference ground state state $|0,0\rangle_{\rm ref}$ are defined by 
\be
\tilde a_1 \tilde a_2\,|0,0\rangle_{\rm target}=0,~~~~a_1 a_2\,|0,0\rangle_{\rm ref}=0
\ee
where 
\be
\tilde a_1&=&\sqrt{\tilde\omega_1\over 2}\,\tilde x_1-i{1\over \sqrt{2\tilde\omega_1}}\tilde p_1,~~~~\tilde a_2=\sqrt{\tilde\omega_2\over 2}\,\tilde x_2-i{1\over \sqrt{2\tilde\omega_2}}\tilde p_2\\
a_1&=&\sqrt{\omega_0\over 2}\,x_1-i{1\over \sqrt{2\omega_0}}p_1, ~~~~a_2=\sqrt{\omega_0\over 2}\,x_2-i{1\over \sqrt{2\omega_0}}p_2~~~\label{axp}
\ee
The relations  between these operators are:
\be
 \tilde a_1&=&{1\over 2\sqrt2}\Big((\alpha+\alpha^{-1})(a_2+a_1)+(\alpha-\alpha^{-1})(a_2^\dag+a_1^\dag)\Big)~~~\label{relation1}\\
 \tilde a_2&=&{1\over 2\sqrt2}\Big((\gamma+\gamma^{-1})(-a_2+a_1)+(\gamma-\gamma^{-1})(-a_2^\dag+a_1^\dag)\Big)~~~\label{relation2}
\ee
where
\be
\alpha&=&\sqrt{{\tilde\omega_1\over \omega_0}},~~~~~~~
\gamma=\sqrt{{\tilde \omega_2\over \omega_0}}~~~\label{omega0}
\ee
We see that both of operators $\tilde a_1$ and $\tilde a_2$ depend on the four kinds operators $a_1^\dag, a_1,a_2^\dag,a_2$.  Thus to consider the  transformation between target state operators and reference state operators we have to consider   4 by 4  matrix in below
\be
\left(\ba{c}
\tilde a_1^\dag\\
\tilde a_2^\dag\\
\tilde a_1\\
\tilde a_2
\ea\right)&=&{\rm \tilde M^{(0)}_{op}}
\left(\ba{c}
a_1^\dag\\
a_2^\dag\\
a_1\\
a_2
\ea\right)~~~~~~~\label{begineq}
\ee
where
\be
{\rm \tilde M^{(0)}_{op}}
&=&{1\over 2\sqrt2}
\left(\ba{cccc}
\alpha+\alpha^{-1}&\alpha+\alpha^{-1}&\alpha-\alpha^{-1}&\alpha-\alpha^{-1}\\
\gamma+\gamma^{-1}&-(\gamma+\gamma^{-1})&\gamma-\gamma^{-1}&-(\gamma-\gamma^{-1})\\
\alpha-\alpha^{-1}&\alpha-\alpha^{-1}&\alpha+\alpha^{-1}&\alpha+\alpha^{-1}\\
\gamma-\gamma^{-1}&-(\gamma-\gamma^{-1})&\gamma+\gamma^{-1}&-(\gamma+\gamma^{-1})
\ea\right)~~~~~~~\label{beginM}
\ee
is the  gate matrix of the ground state in operator approach. 

 Through a SO(4) transformation by  a matrix
\be
A_{SO(4)}=\left(\ba{cccc}
{1\over\sqrt2}&0&{1\over\sqrt2}&0\\
0&{1\over\sqrt2}&0&{1\over\sqrt2}\\
{1\over\sqrt2}&0&-{1\over\sqrt2}&0\\
0&{1\over\sqrt2}&0&-{1\over\sqrt2}\\
\ea\right)~~~~~\label{O(4)}
\ee
the gate matrix $M^{(0)}_{op} $ becomes  a block form, i.e.  ${\rm \tilde M^{(0)}_{op}}\rightarrow {\rm  M^{(0)}_{op}}\equiv A_{SO(4)}^T{\rm \tilde M^{(0)}_{op}} A_{SO(4)}$ where
\be
&&{\rm M^{(0)}_{op}}=\left(\ba{cc}
M_1&0\\
0&M_2\\
\ea\right),~~~{\rm M_1}={1\over\sqrt2}\,
\left(\ba{cc}
\alpha&\alpha\\
\gamma&-\gamma
\ea\right),~~~{\rm M_2}=
{1\over\sqrt2}\,
\left(\ba{cc}
\alpha^{-1}&\alpha^{-1}\\
\gamma^{-1}&-\gamma^{-1}
\ea\right)
\ee
and 
\be
&&\left(\ba{c}
\tilde a_1^\dag +\tilde a_1\\
\tilde a_2^\dag +\tilde a_2\\
\tilde a_1^\dag -\tilde a_1\\
\tilde a_2^\dag -\tilde a_2
\ea\right)
={\rm  M^{(0)}_{op}}
\left(\ba{c}
a_1^\dag +a_1\\
a_2^\dag +a_2\\
a_1^\dag -a_1\\
a_2^\dag -a_2
\ea\right)~~~~~\label{oldTrans1}
\ee
Now the gate matrix  becomes $GL(2,R)\times GL(2,R)$ and we can use the $GL(2,R)$ matrix representation in (\ref{U}) to parameter the  matrix ${\rm M_1}$ and ${\rm M_2}$. 

  For the matrix ${\rm M_1}$  we find that
\be
y(1)&=&{1\over 4}\ln{\tilde \omega_1\tilde \omega_2\over \omega_0^2},~~~~~\rho(1)={1\over 4}\ln{\tilde \omega_2\over \tilde \omega_1}
\ee
It is interesting to see that the values are  exact those in (\ref{trho}) despite $\rm M_w\ne M_1$. Thus the geodesic length is just that in wavefunction approach. 

  For the matrix ${\rm M_2}$
\be
y(1)&=&-{1\over 4}\ln{\tilde \omega_1\tilde \omega_2\over \omega_0^2},~~~~~\rho(1)=-{1\over 4}\ln{\tilde \omega_2\over \tilde \omega_1}
\ee
which gives the same  geodesic length as that from ${\rm M_1}$.

At first sight, the total geodesic length from $GL(2,R)\times GL(2,R)$  will be double comparing to that in wavefunction approach. In fact,  to consider the transformation in the operator approach, for completeness, we had used four operators : $ \tilde  a^\dag_1, \tilde a^\dag_2, \tilde a_1$, and $\tilde a_2$,  not just two operators : $ \tilde a^\dag_1$ and $\tilde a^\dag_2$ which create the target state.  Thus, to count the proper number of the quantum gate we have to reduce the summation by ${1\over 2}$ and
\be
D_{\rm ground~state}^2={1\over 2}\Big(D_{M_1}^2+D_{M_2}^2\Big)={1\over 4}\left(\left(\ln {\tilde \omega_1\over \omega_0}\right)^2+\left(\ln{\tilde \omega_2\over \omega_0}\right)^2\right)=\left(\ln \alpha\right)^2+\left(\ln\gamma\right)^2
\ee
In this way the complexity from operator approach fits to that from wavefunction approach  in (\ref{Distance}).  Note that the SO(4) transformation matrix $A_{SO(4)}$ in (\ref{O(4)}) is just the rotation in the manifold of group $GL(2,R)\times GL(2,R)$  and does not change the geodesic length. 
\subsection{Simple Scheme} 
While  above  algorithm is reasonable we will improve it to a simple scheme which is used to evaluate the complexity of any excited states $|n,m\rangle$ in next section.    First we make  linear combination of  the equation (\ref{oldTrans1}) and transform it to  
\be
&&\left(\ba{c}
A_1^+\\
A_1^-\\
A_2^+\\
A_2^-
\ea\right)={\rm M_{op}}\left(\ba{c}
X\\
Y\\
W\\
Z
\ea\right)
=\left(\ba{cccc}
\alpha&0&0&0\\
0&\alpha^{-1}&0&0\\
0&0&\gamma&0\\
0&0&0&\gamma^{-1}
\ea\right)
\left(\ba{c}
X\\
Y\\
Z\\
W
\ea\right)~~~~~\label{Master}
\ee
where
\be
A_1^+&=&\tilde a_1^\dag +\tilde a_1,~~~~A_1^-=\tilde a_1^\dag -\tilde a_1,~~~~A_2^+=\tilde a_2^\dag +\tilde a_2,~~~~A_2^-=\tilde a_2^\dag -\tilde a_2\\
X&=&{1\over \sqrt 2}\left((a_1^\dag +a_1)+(a_2^\dag +a_2)\right),~~
Y={1\over \sqrt 2}\left((a_1^\dag -a_1)+(a_2^\dag -a_2)\right),\\
Z&=&{1\over \sqrt 2}\left((a_1^\dag +a_1)-(a_2^\dag +a_2)\right)~~
W={1\over \sqrt 2}\left((a_1^\dag -a_1)-(a_2^\dag -a_2)\right)
\ee
In this representation  $(A_1^\pm,~A_2^\pm)$ are the four target  operators and $(X,Y,Z,W)$ are four reference operators.  The gate matrix ${\rm  M_{op}}$ is diagonal  with element  $(\alpha,\alpha^{-1},\gamma,\gamma^{-1})$.  We emphasize that the diagonal property of the gate matrix in this representation  plays the crucial role in calculating the complexity in nth excited state. 

 Considering first the ground state system in which  the target state is defined by $\tilde a_1\tilde a_2|0,0\rangle_{\rm target}=0$.  Although it needs only annihilation  operator here  we need also the associated creation operator to form the basic state (operator) in studying the matrix transformation between referent state and target state,  as that mentioned before.  This means that in the new representation  we need both of operators $A_1^+,A_1^-$, which relate to $a_1^\dag,a_1$,  and  need both of operators $A_2^+,A_2^-$, which relate to $a_2^\dag,a_2$. Therefore the basic operator to represent~ $a_1^\dag\tilde a_2^\dag$~ is ~$(A_1^+,A_1^-)\bigotimes (A_2^+,A_2^-)$.

 Now
\be
(A_1^+,A_1^-)\bigotimes (A_2^+,A_2^-)
&=&\left(\ba{c}
A_1^+A_2^+\\
A_1^+A_2^-\\
A_1^-A_2^+\\
A_1^-A_2^-\\
\ea\right)
=\left(\ba{cccc}
\alpha\gamma&0&0&0\\
0&\alpha\gamma^{-1}&0&0\\
0&0&\alpha^{-1}\gamma&0\\
0&0&0&\alpha^{-1}\gamma^{-1}
\ea\right)
\left(\ba{c}
XZ\\
XW\\
YZ\\
YW
\ea\right)~~~~~\label{(1,1)}
\ee
We see that above gate matrix  is  diagonal and we can  follow the formula (\ref{MD}) to evaluate the associated geodesic length.  The result is
\be
D_{\rm \rm (0,0)}^2&=&{1\over 4}\cdot\left(\left(\ln (\alpha\gamma)\right)^2+\left(\ln  (\alpha\gamma^{-1})\right)^2+\left(\ln  (\alpha^{-1}\gamma)\right)^2+\left(\ln (\alpha^{-1}\gamma^{-1}) \right)^2\right)~~\label{1/2}\\
&=&\left(\ln \alpha\right)^2+\left(\ln \gamma\right)^2
\ee
In this way the complexity calculated in the operator approach fits to the wavefunction approach in (\ref{Distance}).  Before close this section we make two comments about the simple scheme.

1. {Normalization factor} : Note that we add a normalization factor ${1\over 4}$  in (\ref{1/2}).   This is because that to find the gate matrix transformation in the operator approach we use two operators : $ A^+_1,A^-_1$ to relate to $\tilde  a_1$ and use $ A^+_2,A^-_2$ to relate to  $\tilde  a_2$.   Thus the vector space $(A_1^+,A_1^-)\bigotimes (A_2^+,A_2^-)$, which has 4 elements,  is four times the original space with  operators $\tilde  a_1\,\tilde  a_2$, which has only one element.  

 When extending our scheme to the general state $|n,m\rangle_{\rm target}$ which satisfies the relation
\be
\tilde a_1^{n+1}\, \tilde a_2^{m+1}\,|n,m\rangle_{\rm target}=0,~~~~~n,m\ge 0
\ee
it is easy to see that  we have to add a normalization factor 
\be
{Normalization~factor~of~}\,|n,m\rangle_{\rm target}={1\over 2^{n+1}2^{m+1}}~~~~~~\label{factor}
\ee
 to obtain the correct value of complexity in state $|n,m\rangle_{\rm target}$.  In below we always add this normalization factor to calculate the complexity in the excited states. 

2. {Reference state} : Note that in wavefunction approach the target state is the interacting harmonic oscillators which, in terms of norm mode become the state $|0,0\rangle_{\rm target}$ with  frequencies  $\tilde \omega_1$ and  $\tilde \omega_2$.  The reference state is $|0,0\rangle_{\rm ref}$ which is two non-interacting harmonic oscillators with same frequency $\omega_0$.  Since that $\tilde a_1\tilde a_2|0,0\rangle_{\rm target}=0$ while $a_1a_2|0,0\rangle_{\rm ref}= 0$ we see that the target operator is $\tilde a_1,\,\tilde a_2$ while  the reference operator is $ a_1, \,a_2$ in the operator approach.   If turning off the interaction then $\alpha={\tilde \omega_1\over \tilde \omega_0}=\gamma={\tilde \omega_2\over \tilde \omega_0}=1$ and the relations in (\ref{relation1}) and (\ref{relation2}) become
\be
 \tilde a_1&\overset{\alpha=\gamma=1}{\longrightarrow}&{1\over \sqrt2}(a_1+a_2)\not\approx a_1\\
 \tilde a_2&\overset{\alpha=\gamma=1}{\longrightarrow}&{1\over \sqrt2}(a_1-a_2)\not\approx a_2
\ee
Using (\ref{axp}) we see that $(a_1\pm a_2)$  can be regarded as the annihilation operators  at position $x_1\pm x_2$. This means that the reference state is 2 non-interacting  harmonic oscillators states $|0,0\rangle_{\rm ref}$ in which the first and second oscillators are  at position $x_1+x_2$ and $x_1-x_2$ respectively.  After  turn on the interaction the target state is 2 interacting  harmonic oscillators $|0,0\rangle_{\rm target}$.   Thus, for the target state $|n,m\rangle_{\rm target}$, which  will be studied in  below, the corresponding reference state is $|n,m\rangle_{\rm ref}$.

\section{Complexity in Excited States of  2 Harmonic Oscillators}
\subsection{Examples}
{To study the  case of excited states let us begin with the case of ground state. \\
\\
$\bullet$ Ground state $|0,0\rangle_{\rm target}$ :

Ground state is defined by  $\tilde a_1 \,\tilde a_2 |0,0\rangle_{\rm target}=0$. From the argument in previous section, i.e. the  relations $\tilde a_1\sim A_1^++A_1^-$ and $\tilde a_2\sim A_2^++A_2^-$,  we see that to find the gate matrix we have to consider the following relation
\be
(A_1^++A_1^-)(A_2^++A_2^-)&=&(\alpha X +\alpha^{-1} Y)( \gamma Z+\gamma^{-1}W)\nn\\
&=&\alpha\gamma \,XZ+\alpha\gamma^{-1} \,XW+\alpha^{-1}\gamma \, YZ+\alpha^{-1}\gamma^{-1} \,YW
\ee
We see that the coefficients  in last relation, i.e.  $\alpha\gamma,\alpha\gamma^{-1},\alpha^{-1}\gamma,\alpha^{-1}\gamma^{-1}$, are just those in (\ref{1/2}) which are used to calculate the complexity.  
\\
\\
$\bullet$ Excited state  $|1,0\rangle_{\rm target}$ :

For the excited state defined by $(\tilde a_1)^2\, \tilde a_2 |1,0\rangle_{\rm target}=0$  we have to  consider 
\be
&&(A_1^++A_1^-)^2(A_2^++A_2^-)=(\alpha X +\alpha^{-1}Y)^2(\gamma Z+\gamma^{-1}W) =\alpha^2\gamma \,X^2Z \nn \\ 
&&+\alpha^2\gamma^{-1} \,X^2W+2\gamma\,XYZ+2\gamma^{-1}\,XYW+\alpha^{-2}\gamma \,YW+\alpha^{-2}\gamma^{-1} \, Y^2W
\ee
The coefficients  in  the last relation are used to calculate the complexity. 
\\
\\
$\bullet$ Excited state  $|2,0\rangle_{\rm target}$ :

For  the excited state defined by $(\tilde a_1)^3 \,\tilde a_2 |2,0\rangle_{\rm target}=0$  we have to  consider 
\be
&&(A_1^++A_1^-)^3(A_2^++A_2^-)=(\alpha  +\alpha^{-1})^3(\gamma  +\gamma^{-1})\nn\\
&=&\alpha^3\,\gamma+3\alpha\,\gamma+3\alpha^{-1}\,\gamma+\alpha^{-3}\,\gamma+\alpha^3\,\gamma^{-1}+3\alpha\,\gamma^{-1}+3\alpha^{-1}\,\gamma^{-1}+\alpha^{-3}\,\gamma^{-1}
\ee
in which we let $X=Y=Z=W=1$ for simplicity. The coefficients  in  the last relation are used to calculate the complexity. 
\\
\\
$\bullet$ Excited state  $|1,1\rangle_{\rm target}$ :

For  the excited state defined by $(\tilde a_1)^2 \,(\tilde a_2)^2 |1,1\rangle_{\rm target}=0$  we have to  consider 
\be
&&(A_1^++A_1^-)^2(A_2^++A_2^-)^2=(\alpha  +\alpha^{-1})^2(\gamma  +\gamma^{-1})^2\nn\\
&=&\alpha^2\,\gamma^2+\alpha^2\,\gamma^{-2}+\alpha^{-2}\,\gamma^2+\alpha^{-2}\,\gamma^{-2}+2\alpha^2+2\gamma^2+2\alpha^{-2}+2\gamma^{-2}+4
\ee
The coefficients  in  the last relation are used to calculate the complexity.  Note that the constant term does not depend on $\alpha,\gamma$ and contribute null to geodesic.
\\

 Using above coefficients in each excited state, which are associated to the matrix  element of quantum gate, we can  calculate the geodesic length  through  a simple replacement rule
\be
c\,\cdot\,\alpha^{\rm k}  \gamma^{\rm m}\rightarrow c\,\cdot\,\Big[{\rm k}\ln\alpha+ {\rm  m}\ln\gamma \Big]^2~~~\label{simplerule}
\ee
for any constant value of  $c$ \footnote{Constant $c$ represents the multiplicity in which $\alpha^{\rm k}  \gamma^{\rm m}$ appears. It is the coefficient $C^n_{ijk\ell}$ in (\ref{C})}. This replacement rule  has been used in previous sections and is checked in (\ref{MD}).   Then we can quickly find that 
\be
D_{\rm (0,0)}^2&=&\left(\ln \alpha\right)^2+\left(\ln \gamma\right)^2\\
D_{\rm (1,0)}^2&=&2\left(\ln \alpha\right)^2+\left(\ln \gamma\right)^2\\
D_{\rm (2,0)}^2&=&3\left(\ln \alpha\right)^2+\left(\ln \gamma\right)^2\\
D_{\rm (1,1)}^2&=&2\left(\ln \alpha\right)^2+2\left(\ln \gamma\right)^2
\ee
after imposing the proper normalization factor ${1\over 2^{n+1}2^{m+1}}$ mentioned in (\ref{factor}).

 We can easily extend the calculations to higher excited states.   The results indicate that the square of length of the quantum circuit in general  state $|n,m\rangle_{\rm target}$ is $D_{\rm (n,m)}^2=(n+1)\left(\ln \alpha\right)^2+(m+1)\left(\ln \gamma\right)^2$. We show this formula  in below.
\subsection{General Formula}
Using previous arguments the property of the gate matrix in the state $|n-1,m-1\rangle_{\rm target}$ is read from the coefficients in the following  expansion
\be
(\alpha+\alpha^{-1})^n(\gamma+\gamma^{-1})^m&=&\sum^{n,m}_{i,j,k,\ell\ge0}\,(\alpha)^i\, (\alpha^{-1})^j\,(\gamma)^k\,(\gamma^{-1})^\ell\,\,C^{n,m}_{ijk\ell}\nn\\
&=&\sum^{n,m}_{i,j,k,\ell\ge0}\,\alpha^{i-j}\,\gamma^{k-\ell}\,\,C^{n,m}_{ijk\ell},~~~~~n,m\ge1~~~\label{C}
\ee
where the summation is constrained by $i+j=n,~k+\ell=m$.  $C^{n,m}_{ijk\ell}$ is the coefficients in the expansion. To calculate the geodesic length from above coefficients  we can use the replacement rule  in (\ref{simplerule}). Thus
\be
D_{\rm (n-1,m-1)}^2&=&{1\over 2^n2^m}\sum^{n,m}_{i,j,k,\ell\ge0}\, \left[ (i-j)\ln\alpha + (k-\ell)\ln\gamma \right]^2\,\,C^{n,m}_{ijk\ell}\nn\\
&=&{1\over 2^n2^m}\left(\sum^{n,m}_{i,j,k,\ell\ge0}\, (i-j)^2\,C^{n,m}_{ijk\ell}\right)\Big(\ln\alpha \Big)^2+{1\over 2^n2^m}\left(\sum^{n,m}_{i,j,k,\ell\ge0}\, (k-\ell)^2\,C^{n,m}_{ijk\ell}\right)\Big(\ln\gamma \Big)^2~~~~~~\nn\\
&=&n\left(\ln \alpha\right)^2+m\left(\ln \gamma\right)^2,~~~n,m\ge1~\label{main1}
\ee
where we have used the summation formulas in (\ref{sum1}) and (\ref{sum2}). 
\subsection{Wavefunction of Excited States}
In this subsection we will make a comment about the wavefunction of the excited state in this paper.  For the simple harmonic oscillator we have the well-known properties
\be
H=\hbar\omega\,(a^\dag_\omega a_\omega),~~~~H|n_\omega\rangle=\hbar\omega\,\Big(n_\omega+{1\over 2}\Big)|n_\omega\rangle,~~~~~~~|n_\omega\rangle={(a^\dag_\omega)^{n_\omega}\over \sqrt{n_\omega\,!}}\,|0\rangle
\ee
Therefore, while  $|0\rangle$ is the ground (vacuum) state we can regard $|n_\omega\rangle$ as the nth excited state. On the other hand, in occupation number representation the state $|n_\omega\rangle$ can also be represented, for example,  as
\be
|n_\omega\rangle=|\overbrace{1_\omega,....,1_\omega}^{n_\omega}\,\rangle~~~~~~\label{11excite}
\ee
Consider, for example, the excited  state  $|2_{\omega_a},1_{\omega_b}\rangle$ which can be regarded as the system with two oscillators, N=2.  One of which is the second excited state ($\Psi^{\rm (2th)}_{\omega_a}$) with frequency $\omega_a$ and another is the  first  excited state ($\Psi^{\rm (1th)}_{\omega_b}$) with frequency $\omega_b$.  The corresponding wavefunction in the first quantization is 
\be
{1\over \sqrt2}\Big(\Psi^{\rm (2th)}_{\omega_a}(x_1)\Psi^{\rm (1th)}_{\omega_b}(x_2)+\Psi^{\rm (1th)}_{\omega_b}(x_1)\Psi^{\rm (2th)}_{\omega_a}(x_2)\Big)~~~~~\label{wave1}
\ee
On the other hand, using the relation (\ref{11excite}) we find another representation $|2_{\omega_a},1_{\omega_a}\rangle=|1_{\omega_a},1_{\omega_a},1_{\omega_b}\rangle$  and corresponding wavefunction in the first quantization is 
\be
{1\over \sqrt3}\Big(\Psi^{\rm (1th)}_{\omega_a}(x_1)\Psi^{\rm (1th)}_{\omega_a}(x_2)\Psi^{\rm (1th)}_{\omega_b}(x_3)+\Psi^{\rm (1th)}_{\omega_a}(x_1)\Psi^{\rm (1th)}_{\omega_b}(x_2)\Psi^{\rm (1th)}_{\omega_a}(x_3)+\Psi^{\rm (1th)}_{\omega_b}(x_1)\Psi^{\rm (1th)}_{\omega_a}(x_2)\Psi^{\rm (1th)}_{\omega_a}(x_3)\Big)\nn\\
~~~~~~~\label{wave2}
\ee 
In this representation  there are three oscillators, N=3, and two among them  have the same frequency $\omega_a$.

Now, both of the N=2 wavefunction (\ref{wave1}) and  the N=3 wavefunction (\ref{wave2}) could be used to represent state $|2_{\omega_a}, 1_{\omega_b} \rangle$ and,  at first sight,  there is ambiguity in  choosing a wavefunction to represent $|2_{\omega_a}, 1_{\omega_b} \rangle$ in our calculation.  In fact, from (\ref{frequency}) we see that the frequencies  of  the two oscillators, N=2, is different from the frequencies  of the three oscillators, N=3. Therefore, after using (\ref{frequency}) to choose the proper frequencies, $(\omega_a,\omega_b)$, of the  N=2 oscillators the wavefunction in (\ref{wave1}) can be used to represent the state $|2_{\omega_a}, 1_{\omega_b} \rangle$. In this case,  the N=3 wavefunction  in (\ref{wave2}) could not  be used to represent the state $|2_{\omega_a}, 1_{\omega_b} \rangle$ since the frequencies $(\omega_a,\omega_a,\omega_b)$ do not fit the relation (\ref{frequency}).

Therefore,  the state $|n_{\omega_a},m_{\omega_b}\rangle$ considered in this paper is the system  with 2 oscillators in which one is at nth excited state with frequency $\omega_a$ while another is at mth excited state with frequency $\omega_b$. The corresponding wavefunction in the first quantization  is
\be
|n_{\omega_a},m_{\omega_b}\rangle\rightarrow {1\over \sqrt2}\Big(\Psi^{\rm (nth)}_{\omega_a}(x_1) \Psi^{\rm (mth)}_{\omega_b}(x_2)+\Psi^{\rm (mth)}_{\omega_b}(x_1)\Psi^{\rm (nth)}_{\omega_a}(x_2) \Big)
\ee
The arguments can be applied to many oscillators system.

\section{Complexity in N Harmonic Oscillators  and Lattice Scalar Field Theory}
{\color{black}Above calculations are performed on the two coupled harmonic oscillators and, for self-consistent, in this section we will describe how these calculations can be extended to scalar field theory. 

The lattice version of one-dimensional free scalar field can be described by one dimensional N harmonic oscillators.  The Hamiltonian  is  that extended  two coupled harmonic oscillator in (\ref{LH1}) to N harmonic oscillators
\be
H={1\over2}\sum_{k=1}^N\,p_k^2+\omega^2 x_k^2+\Omega^2(x_k-x_{k+1})^2
\ee
with periodic boundary condition $x_{\rm k+N+1}=x_k$  \cite{Jefferson1707}. In the normal coordinate the Hamiltonian becomes 
\be
H={1\over2}\sum_{k=1}^N\,\tilde p_k^2+\tilde\omega_k^2 \tilde x_k^2
\ee
where \cite{Jefferson1707}
\be
\tilde x_k&=& {1\over \sqrt N}\sum_{j=1}^N\,exp\Big({-2\pi i k\over N} \,j\Big)\,x_j\\
\tilde p_k&=&{1\over \sqrt N}\sum_{j=1}^N\,exp\Big({-2\pi i k\over N} \,j\Big)\,p_j \\
\tilde \omega_k^2&=& \omega^2+4\Omega^2\,\sin^2{\pi k\over N}~~~~\label{frequency}
\ee
In the operator approach we need the relations between the annihilation  operators in original coordinates and in normal coordinate. The conventional definitions  are
\be
\tilde a_k&=&\sqrt{\tilde\omega_k\over 2}\,x_k-i{1\over \sqrt{2\tilde\omega_k}}p_k,~~~~
a_k=\sqrt{\omega_0\over 2}\,x_k-i{1\over \sqrt{2\omega_0}}p_k
\ee
in which $\omega_0$ is the reference state frequency used in  (\ref{omega0}).  The relation  between the operators is 
\be
\tilde a_k&=&{1\over 2\sqrt N}\Big(\alpha_k+\alpha_k^{-1}\Big)\Big(\sum_{j=1}^N\,exp\Big({-2\pi i k\over N} \,j\Big)\,a_j\Big)+{1\over 2\sqrt N}\Big(\alpha_k-\alpha_k^{-1}\Big)\Big(\sum_{j=1}^N\,exp\Big({-2\pi i k\over N} \,j\Big)\,a_j^\dag \Big)\nn\\
\ee
where
\be
\alpha_k&=&\sqrt{{\tilde\omega_k\over \omega_0}},~~k=1,\cdot\cdot\cdot,N
\ee
Following the investigations in previous sections we can now use the following reference operators $\{X^+_k,X_k^-\}$ and target operators $\{A^+_k,A_k^-\}$
\be
A^+_k&\equiv&\tilde a^\dag_k+\tilde a_k=\alpha_k\,\left({1\over \sqrt N}\,\sum_{j=1}^N\,exp\Big({-2\pi i k\over N} \,j\Big)\,(a_j^\dag+a_j)\right)\equiv\alpha_k\,X_k^+\\
A^-_k&\equiv&\tilde a^\dag_k-\tilde a_k=\alpha_k^{-1}\,\left({1\over \sqrt N}\,\sum_{j=1}^N\,exp\Big({-2\pi i k\over N} \,j\Big)\,(a_j^\dag-a_j)\right)\equiv\alpha_k^{-1}\,X_k^-
\ee
In this representation the gate matrix is diagonal.

Now we can follow the investigation in the previous section  to calculate the complexity of state $|n_1 \cdot \cdot \cdot n_N\rangle$ in lattice scalar field theory.  The extension of (\ref{C}) tells us that we  shall consider the following expansion 
\be
\prod_{k=1}^N\,(\alpha_{k}+\alpha_{k}^{-1})^{n_k}&=&\sum^{\{n_k\}}_{\{i_k\ge0,\,j_k\ge0\}}\,\prod_{k=1}^N\,(\alpha_{k})^{i_k}\, (\alpha_{k}^{-1})^{j_k}\,\,C^{\{n_k\}}_{\{i_k,j_k\}}\nn\\
&=&\sum^{\{n_k\}}_{\{i_k\ge0,\,j_k\ge0\}}\,\prod_{k=1}^N\,\,(\alpha_{k})^{i_k-j_k}\,\,C^{\{n_k\}}_{\{i_k,j_k\}}~~~\label{CC}
\ee
where the summation is constrained by $i_k+j_k=n_k$.  $C^{\{n_k\}}_{\{i_k,j_k\}}$ is the coefficients in expansion. To calculate the geodesic length from above coefficients  we can use the replacement rule  in (\ref{simplerule}). Thus
\be
D_{\rm (\{n_k-1\})}^2&=&{1\over 2^{n_1+\cdot\cdot\cdot+n_N}}\sum^n_{\{i_k\ge0,\,j_k\ge0\}}\,\Big[\sum_{k=1}^N\,(i_k-j_k)\,\ln (\alpha_{k})\Big]^2\,\,C^{\{n_k\}}_{\{i_k,j_k\}}\nn\\
&=&{1\over 2^{n_1+\cdot\cdot\cdot+n_N}}\sum_{k=1}^N\,\Big(\sum^{\{n_k\}}_{\{i_k\ge0,\,j_k\ge0\}}\,(i_k-j_k)^2\,\,C^{\{n_k\}}_{\{i_k,j_k\}}\Big)\,\ln (\alpha_{k})\\
&=&\sum_{k=1}^N\,n_k\,\ln (\alpha_{k})~~~\label{main2}
\ee
where we have used the summation formula in (\ref{sumN}). Above result is the extension of (\ref{main1}) which is the case of N=2.

The extension to d-1 dimensional coupled harmonic oscillation (which describes the d dimensional free scalar field theory) is just to replace above $k$ by $\vec k=(k_{1},k_{2},\cdot\cdot\cdot,k_{d-1})$, as that described in \cite{Jefferson1707}.} The result is
\be
D_{\rm (\{n_{k_i}-1\})}^2&=&\sum_{i=1}^{d-1}\,\sum_{k=1}^N\,n_{k_i}\,\ln (\alpha_{k_{i}}),~~~~~\alpha_{k_{i}}=\sqrt{{\tilde\omega_{k_i}\over \omega_0}}~~~~\label{final}
\ee
which is the  excited-state complexity of the d dimensional free scalar field theory on lattice.
\section{Discussions}
In this paper we present the operator approach to calculate the complexity of free scalar field.  We first  quickly review the method in  \cite{Jefferson1707} and  describe the lattice free scalar field as coupled harmonic oscillators.  Then, as a first step,  we investigate  the 2 coupled oscillators and regard  the transformation  between the creation operators  in the reference state and target state as the gate matrix. We find that the gate matrix  can be reduced to diagonal form after orthogonal transformation,  then we calculate the complexity of the ground state. The result reproduces the known value of ground-state complexity in \cite{Jefferson1707}.  While  the excited-state wavefunction of harmonic oscillation  is not pure  exponential form and  the wavefunction approach is hard to work  the  operator approach, which need not to use the explicit form of the wave function,  can be used to study the complexity in the excited state.  We explicitly calculate the complexity in several excited states  and  present a simple derivation to find the general formula of  complexity in the any excited states.  Finally, we describe how the calculations in the two coupled harmonic oscillators  can be extended to N coupled harmonic oscillators which describes the lattice free scalar field theory.

Some more studies are necessary  to clarify the property of using the operator approach to calculate the complexity:

1.  We have  calculated the complexity of  excited states from field theory.    It is  interesting to find the holographic  results to compare our results.  Note that, as that discussed in section 3.2, the reference state of  ground state is different from that of excited states and are different ones for each excited state. This 
is problematic for the holographic motivation of this work.  For example, should the growing size of a black hole reflect the relative complexity between a time-dependent microstate and a time-dependent reference? If we allow the reference to be time-dependent, then we are essentially introducing a time-dependent free parameter into the problem of explaining one time-dependent quantity (the volume or WdW patch action, or whatever), which means we have zero predictability and zero falsifiability.

2.  We only study the free scalar field in this paper. The extension to interacting theory and fermion theory is deserved to  study.

3.  Since that the operator approach needs not know the position-space wavefunction it can be used to study the complexity in the spin system which appears in many models of condense matter.

 These problems are under studied.
\\
\\
{\bf Acknowledgments} :  The author is grateful to the anonymous referee for many useful comments which have significantly improved the quality of paper after several corrections.
\\
\\
\appendix
\section{Geodesic Length in Equivalent Gate Matrix}
Consider  following two  gate matrix
\be 
M_1=\left(\ba{cc}\alpha&\alpha\\ \gamma&-\gamma \ea\right)~~~;~~~M_2= \left(\ba{cc} \alpha&0\\ 0&\gamma \ea\right)
\ee
where parameters $\alpha$ and $\gamma$ could be any values. Since that 
\be
M_1\cdot  \left(\ba{c}\Psi_a\\ \Psi_b\ea\right)=M_2\cdot  \left(\ba{c}\Psi_a+\Psi_b\\ \Psi_a-\Psi_b\ea\right)
\ee
we say that  matrix  $M_2$  is the matrix $M_1$  after rearrangement (or linear combination).  \\

 We can regard  both matrix  as GL(2,R) and use the function form in  (\ref{GL}) to calculate the geodesic lengths.  Both give the same geodesic length
\be
D^2_{GL(2,R)}=2({1\over2}(\ln\alpha+\ln\gamma))^2+2({1\over2}(\ln\alpha-\ln\gamma)^2=(\ln\alpha)^2+ (\ln\gamma)^2~~~\label{appendex1}
\ee
On the other hand  we can regard the second matrix as $R\times R$ and express it in a general matrix
\be
M_{R\times R}= \left(\ba{cc}e^{y_1}&0\\0&e^y_2\ea\right)~~~~y_1=\ln\alpha,~~y_2=\ln\gamma
\ee
The associated metric  and geodesic length are 
\be
ds_{R\times R}^2&=&{\rm tr}(de^{y_1}\,e^{-y_1})\,{\rm tr}(de^{y_1}\,e^{-y_1})+{\rm tr}(de^{y_2}\,e^{-y_2})\,{\rm tr}(de^{y_2}\,e^{-y_2})=(dy_1)^2+(dy_2)^2\\
D^2_{R\times R}&=&(\ln\alpha)^2+ (\ln\gamma)^2
\ee
which is consistent with (\ref{appendex1}). 

 Therefore for a diagonal gate matrix $M$ the associated geodesic length $D^2_{M}$  is
\be
D^2_{M}=diagonal\Big(e^{y_1},\cdot\cdot\cdot,e^{y_k},\cdot\cdot\cdot,e^{y_n}\Big) ~~\Rightarrow~~~~~D^2_{M}=\sum_{k=1}^n(\ln y_k)^2~~~\label{MD}
\ee
The relation  is extensively used in this paper.
\section{Summation Formula}
In this appendix we consider two summation formulas :\\
$\bullet$ Begin with the definition 
\be
(a+b)^n(c+d)^m&=&\sum^{n,m}_{i,j,k,\ell\ge0}\,a^i\, b^j\,c^k\,d^\ell\,\,C^{n,m}_{ijk\ell}
\ee
where the summation is constrained by $i+j=n,~k+\ell=m$.  $C^{n,m}_{ijk\ell}$ is the coefficients in expansion.   Considering three cases of the derivative of above relation : 1. Derivative one time with respective to $a$.  2. Derivative two times with respective to $a$.  3.  Derivative with respective to $a$ then to $b$.  Let $a=b=c=d=1$ in above three kinds of derivative we have three relations 
\be
n2^{n-1}2^{m}&=&\sum^{n,m}_{i,j,k,\ell\ge0}\,i\,\,C^{n,m}_{ijk\ell}\\
n(n-1)2^{n-1}2^{m}&=&\sum^{n,m}_{i,j,k,\ell\ge0}\,i(i-1)\, \,C^{n,m}_{ijk\ell}\\
n(n-1)2^{n-1}2^{m}&=&\sum^{n,m}_{i,j,k,\ell\ge0}\,ij\,\,C^{n,m}_{ijk\ell}
\ee
Use above relations we find a summation formula
\be
\sum^n_{i,j,k,\ell\ge0}\, (i-j)^2\,C^n_{ijk\ell}=n\,2^n2^m~~\label{sum1}
\ee
In a similar way we can find another formula
\be
\sum^n_{i,j,k,\ell\ge0}\, (k-\ell)^2\,C^n_{ijk\ell}=m\,2^n2^m~~\label{sum2}
\ee
Above two formulas are used to the obtain the general formula (\ref{main1}). 
\\
\\
$\bullet$ Begin with the definition 
\be
(a_1+b_1)^{n_1}\cdot\cdot\cdot(a_k+b_k)^{n_k}\cdot\cdot\cdot(a_N+b_N)^{n_N}&=&\sum^{\{n_s\}}_{\{i_s\ge0,\,j_s\ge0\}}\,\,a_1^{i_1}\,b_1^{j_1}\cdot\cdot\cdot  a_k^{i_k}\,b_k^{j_k} \cdot\cdot\cdot a_N^{i_N}\,b_N^{j_N}\,\,C^{\{n_s\}}_{\{i_s,j_s\}}\nn\\
\ee
where the summation is constrained by $i_s+j_s=n_s$, $s=1,2\cdot\cdot\cdot, N$.  $C^{\{n_s\}}_{\{i_s,j_s\}}$ is the coefficients in expansion. Considering three cases of the derivative of above relation : 1. Derivative one time with respective to $a_k$.  2. Derivative two times with respective to $a_k$.  3.  Derivative with respective to $a_k$ then to $b_k$.  Let $a_1=b_1=\cdot\cdot\cdot=a_N=b_N=1$ after above three kinds of derivative we  have three relations 
\be
n_k\,\times\,2^{-1+\sum_s n_s}&=&\sum^{\{n_s\}}_{\{i_s\ge0,\,j_s\ge0\}}\,\,i_k\,\, C^{\{n_s\}}_{\{i_s,j_s\}}\\
n_k(n_k-1)\,\times\,2^{-2+\sum_s n_s}&=&\sum^{\{n_s\}}_{\{i_s\ge0,\,j_s\ge0\}}\,\,i_k(i_k-1)\,\,C^{\{n_s\}}_{\{i_s,j_s\}}\\
n_k(n_k-1)\,\times\,2^{-2+\sum_s n_s}&=&\sum^{\{n_s\}}_{\{i_s\ge0,\,j_s\ge0\}}\,\,i_k\,j_k\, C^{\{n_s\}}_{\{i_s,j_s\}}
\ee
Use above relations we find a summation formula
\be
\sum^{\{n_s\}}_{\{i_s\ge0,\,j_s\ge0\}}\,\, (i_k-j_k)^2\,C^{\{n_s\}}_{\{i_s,j_s\}}=2^{\sum_s n_s}~\times~n_k~\label{sumN}
\ee
which is used to the obtain the general formula (\ref{main2}).  The case of  $N=2$ above relation reduces to (\ref{sum1}) and (\ref{sum2}) .

\begin{center} {\bf REFERENCES}\end{center}

\end{document}